\documentclass[twocolumn,amsmath,amssymb,aps,prc]{revtex4-2}
\usepackage{graphicx,here}
\usepackage{dcolumn,bm,hyperref,footnote}
\usepackage{float}
\usepackage[dvipsnames,svgnames,x11names]{xcolor}
\usepackage[authormarkup=none,defaultcolor=BrickRed]{changes}
\usepackage{multirow}
\usepackage{lipsum}
\usepackage{tikz}
\usepackage{booktabs}

 \DeclareMathOperator*{\argmax}{argmax}
 \usepackage{bbold}
\usepackage{subcaption}
\usepackage{placeins}
\DeclareMathOperator*{\argmin}{arg\,min} 
\begin{document}

\title{An optical-lensing inspired data thinning method for nuclear cross section data}

\author{M. Imbri{\v s}ak}
\email{marko.imbrisak@gmail.com}
\affiliation{Los Alamos National Laboratory, Los Alamos, NM, 87545, USA}
\author{A. E. Lovell}
\affiliation{Los Alamos National Laboratory, Los Alamos, NM, 87545, USA}
\author{M. R. Mumpower}
\affiliation{Los Alamos National Laboratory, Los Alamos, NM, 87545, USA}

\date{\today}

\begin{abstract}
In the study of nuclear cross sections, the computational demands of data assimilation methods can become prohibitive when dealing with large data sets. We have developed a novel variant of the data thinning algorithm, inspired by the principles of optical lensing, which effectively reduces data volume while preserving critical information. We show how it improves fitting through a toy problem and for several examples of total cross sections for neutron-induced reactions on rare-earth isotopes. We demonstrate how this method can be applied as an efficient pre-processing step prior to smoothing, significantly improving computational efficiency without compromising the quality of uncertainty quantification.
\end{abstract}

\maketitle

\section{Introduction\label{sec:intro}}
{\color{black}
When fitting physical models to experimental data, it is common to encounter large, and often correlated, datasets. This poses challenges not only in terms of computational cost—particularly when the models are themselves expensive to calculate—but also in data assimilation and uncertainty quantification. To address these issues, various data thinning techniques have been developed to reduce the volume of data used during optimization without significantly degrading model accuracy. In this paper, we introduce a novel data thinning method inspired by optical lensing. Such data thinning techniques are broadly applicable to challenges across nuclear and astrophysical modeling. In nuclear physics, evaluated data projects like ENDF/B \citep{chadwick2011endf} must assimilate thousands of correlated cross-section datasets. Surrogate reactions \citep{escher2012surrogate} also require efficient treatment of dense experimental data. In astrophysics, r-process nucleosynthesis \citep{kajino2019rprocess,mumpower2016impact}, supernova modeling \citep{langanke2003weak}, and EOS inference from neutron star mergers \citep{oertel2017eos} are all data-intensive domains where thinning methods can mitigate computational bottlenecks in uncertainty propagation and model calibration.

The Hauser-Feshbach (HF) statistical model is used to compute nuclear reactions on medium- to heavy-mass nuclei. The HF model calculates energy-averaged nuclear reaction cross sections and incorporates a width fluctuation correction for overlapping resonances. Out of the several HF codes available on the market, including EMPIRE \cite{INDC060}, TALYS \cite{Koning2012}, CCONE \cite{Iwamoto2007}, we focus our attention on the LANL-developed CoH$_3$ code \cite{Kawano2010,Kawano2019}. These codes in general handle multi-particle evaporation from a compound nucleus and provide not just the reaction cross sections, but also a wealth of information on other observables such as the energy and angular distributions of secondary particles, $\gamma$-ray production cross sections, and the production of isomeric states. 
The CoH$_3$ code \cite{Kawano2010,Kawano2019} integrates the coupled-channels optical model to achieve precise nuclear reaction calculations within the keV to tens of MeV energy range. The CoH$_3$ package is structured into modules that provide a comprehensive toolkit for nuclear reaction analysis: the one-body potential mean-field theory, the coupled-channels optical model, and the HF statistical decay model, modules that handle direct and semidirect radiative capture, pre-equilibrium processes, and prompt fission neutron emission.

The Experimental Nuclear Reaction Data  (EXFOR) \cite{ENDF2014} database is often used to aggregate data of multiple nuclear reaction data, which includes cross section experiments. The well-known problem in such data aggregation is that it neglects experimental correlations, such as those between experiments \cite{neudecker_advanced_2014}. To account for this, an iterative generalized least-squares algorithm, the Full Bayesian Evaluation Technique (FBET) \cite{leeb_consistent_2008,leeb_geneus_2011} was developed for nuclear data evaluation. FBET is a data assimilation method that evaluates nuclear data at a chosen energy grid and provides an estimate for correlated experimental uncertainties using a linearized Bayesian update procedure. These experimental uncertainties rank equally as important as uncertainties from model defects \cite{neudecker_impact_2013}.
The FBET approach to treating correlated measurements is not unique to nuclear physics and belongs to a wider class of data assimilation methods which  have been successfully applied in geophysical models \cite{stewart_data_2013,cardinali_observation_2013} and industrial applications \cite{cheng_background_2019}. A problem often encountered in these fields is computational complexity of large data sets needed for data assimilation. One of the solutions to this problem is reducing the amount of data in a statistically consistent way by finding an optimal subset of data that describes the wanted quantity equally well as the entire dataset. One class of such methods is data thinning, which accomplishes this using either clustering \cite{ochotta_adaptive_2005} or iteratively applying a selected statistic \cite{Ochotta2005}.

 We have developed a data thinning algorithm inspired by optical lensing to improve the analysis of cross section data in the region with pronounced resonances. The data for many nuclei below $\sim 1\,\mathrm{MeV}$ is abundant, but due to finite-energy sampling, it can often be hard to attribute a data point to a resonance peak versus the ``smooth'' underlying HF curve. While in principle, FBET can handle such data, it is often impractical to apply it to large datasets due to memory requirements. We propose our lensing method as a pre-processing step in order to identify the most impactful data points in a measurement set.
 }

We describe the FBET approach, the CoH$_3$ code, and present the percentile-based and KDE-based data thinning methods in sections \ref{sec:percentile} and \ref{sec:kde}. In Sec. \ref{sec:testing}, we then describe the testing of the lensing method and present our results in \ref{sec:results}.

\section{Lensing}\label{sec:lensing}
\subsection{Percentile-based lensing estimates}\label{sec:percentile}
 Our lensing-inspired thinning method requires binning a dataset into $N_b$ energy bins, $[x_1,x_2],\cdots, [x_{b}, x_{b+1}],\cdots, [x_{N_b}, x_{N_b+1}]$, and computing the percentiles corresponding to the minimum, $p=0$, median, $p=50$, and maximum $p=100$ values of cross sections within the bin $b$, $y_{p,b}$. We then introduce the likelihood:
 \begin{align}
     L_b(y,\sigma_y) &=\prod\limits_{p\in\{0,50,100\}}e^{-\frac{1}{2}\left(\frac{y-y_{p,b}}{\sigma_y}\right)^2},
 \end{align}
 and find the index of the measurement from bin $b$ that maximizes the likelihood of bin $b'$, $L_{b'}$,
\begin{align}
    i_{b,b'} &=\argmax\limits_{x_i\in[x_b,x_{b+1}]}L_{b'}\left(y_i, \sigma_{y_i}\right).
\end{align}
The procedure is as follows. We iterate over bins $b=1,\cdots, N_b$ and find the measurements $i_{b,b+1}$. We thereby get one measurement per bin. We then remove these measurements from the minimization and repeat the procedure to find the measurements $i'_{b,b+1}$. This procedure can be repeated until the desired limit on the number of measurements per bin is reached.

\FloatBarrier

\begin{figure*}
\begin{tikzpicture}[scale=.5, transform shape,block/.style={
      rectangle,
      draw=blue,
      thick,
      fill=blue!20,
      align=center,
      rounded corners,
      minimum height=4cm,scale=1.5
    },block2/.style={
      rectangle,
      draw=blue,
      thick,
      fill=blue!20,
      align=center,
      rounded corners,
      minimum height=2cm
    },block3/.style={
      rectangle,
      draw=blue,
      thick,
      fill=blue!20,
      align=center,
      rounded corners,
      minimum height=2cm
    }]
    \node[block,text width=3cm,align=center] (a) at (0,0)   {Update function 
    and Jacobian values based on current parameters}; 
    \node[block,text width=3cm,align=center] (b) at (6,0)   {Update scaling matrix and damping parameter}; 
    \node[block,text width=3cm,align=center] (c) at (12,0)   {Propose parameter step, evaluate function at new parameters}; 
    \node[block,text width=3cm,align=center] (d) at (18,0)   {Accept or reject parameter step based on cost decrease}; 
    \node[block,text width=3cm,align=center] (e) at (24,0)   {Check convergence criteria}; 
    \node[block,text width=3cm,align=center] (f) at (30,0)   {Stop if criteria met or limits exceeded}; 
    
    \node[block2,text width=4cm,align=center,scale=1.5] (g) at (6,7)   {%
    \begin{align*}
D_{\mu\nu}&=\begin{cases}g _{\mu\nu},& \mu=\nu\\
        0, & \mu\neq \nu\end{cases}\\
g_{\mu\nu}^{(\lambda)}&=g_{\mu\nu}+\lambda D_{\mu\nu}
    \end{align*}
    };
    \node[block3,text width=7cm,align=center,scale=2] (h) at (12,-7)   {%
    \begin{align*}
        \delta\theta^\mu=-g^{(\lambda)}{}^{\mu\nu}\sum\limits_{i=1}^{N} \partial_\nu r^{i} r^{i}
    \end{align*}
    };
    \node[block2,text width=9cm,align=center,scale=1.5] (i) at (18,7)   {%
    \begin{align*}
        \theta^\mu_{n+1}&=\begin{cases}
            \theta^\mu_{n}, &  \chi^2(\theta_n+\delta\theta)>\chi^2(\theta_n)\\
            \theta^\mu_{n}+\delta\theta^\mu, & \chi^2(\theta_n+\delta\theta)< \chi^2(\theta_n)
        \end{cases}\\
        \lambda_{n+1}&=\begin{cases}
            \lambda_{n}\lambda_u, & \chi^2(\theta_n+\delta\theta)> \chi^2(\theta_n)\\
            \frac{\lambda_{n}}{\lambda_d}, & \chi^2(\theta_n+\delta\theta)< \chi^2(\theta_n)
        \end{cases}
    \end{align*}
    };
    \node (j) at (0,-10){};
    \node (k) at (24,-10){};
    \draw[->,thick] (a)--(b);
    \draw[->,thick] (b)--(c);
    \draw[->,thick] (c)--(d);
    \draw[->,thick] (d)--(e);
    \draw[->,thick] (e)--(f);
    \draw[->,thick] (g)--(b);
    \draw[->,thick] (h)--(c);
    \draw[->,thick] (i)--(d);
  \draw[->,thick] (e)|-(j)-|(a);
\end{tikzpicture}  
\caption{Schematic representation of the Levenberg-Marquardt procedure.}\label{fig:Schema}
\end{figure*}
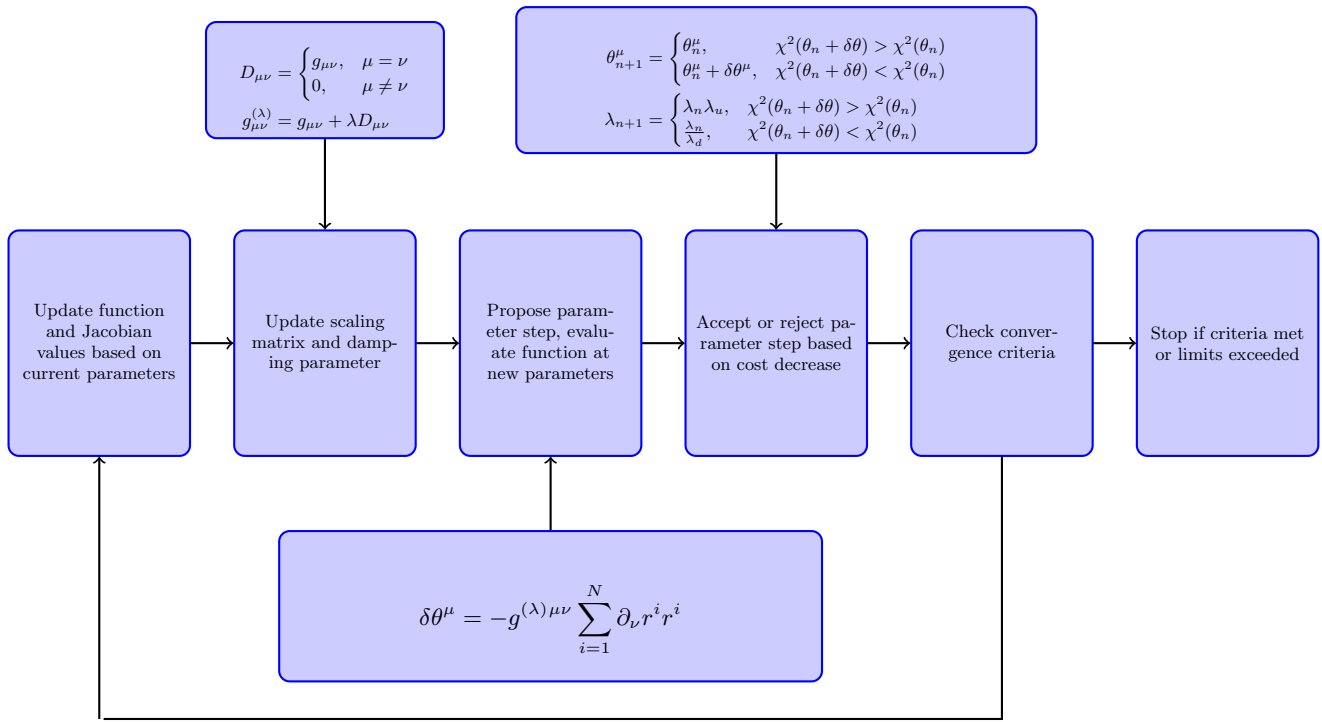

\begin{table*}[t]
    \centering
    
    \resizebox{\textwidth}{!}{%
    \begin{tabular}{llllllllll}
\toprule
Lensing method & slope & intercept & $t(\mathrm{slope})$ & $P(\mathrm{slope})$ & $t(\mathrm{intercept})$ & $P(\mathrm{intercept})$ & Chow's $F$ & $P($Chow's $F)$ & Evaluation time [ms] \\
\midrule
Percentile-based & $0.98\pm 0.31$ & $2.62\pm 0.16$ & 3.33 & 0.02 & 3.85 & 0.01 & 0.12 & 0.9965 & $11.62\pm 0.95$ \\
KDE-based & $1.91\pm 0.09$ & $2.20\pm 0.05$ & 1.02 & 0.35 & 4.07 & 0.01 & 0.18 & 0.9965& $150\pm 30$ \\
KDE$\sigma$-based & $2.02\pm 0.09$ & $2.21\pm 0.06$ & 0.23 & 0.82 & 3.46 & 0.01 & 0.02 & 0.9966 & $148\pm 32$ \\
\bottomrule
\end{tabular}}\caption{Comparison of linear model simulations' statistics. The columns slope and intercept show the best-fitting linear model parameters to a particular lensed dataset. The $t$-labeled columns show their respective $t$-statistics, while $P$-labeled columns show the corresponding 2-sided $P$-values. The $F$ statistic and its corresponding $P$-value was computed to compare the best-fitting values to the best-fitting values of the bootstraps. The median and standard deviation  of a particular lensing method's computation time were estimated from bootstrap simulations' computation times.}\label{tab:bootstrap}
   

\end{table*}
\begin{figure*}
\includegraphics[width=\linewidth]{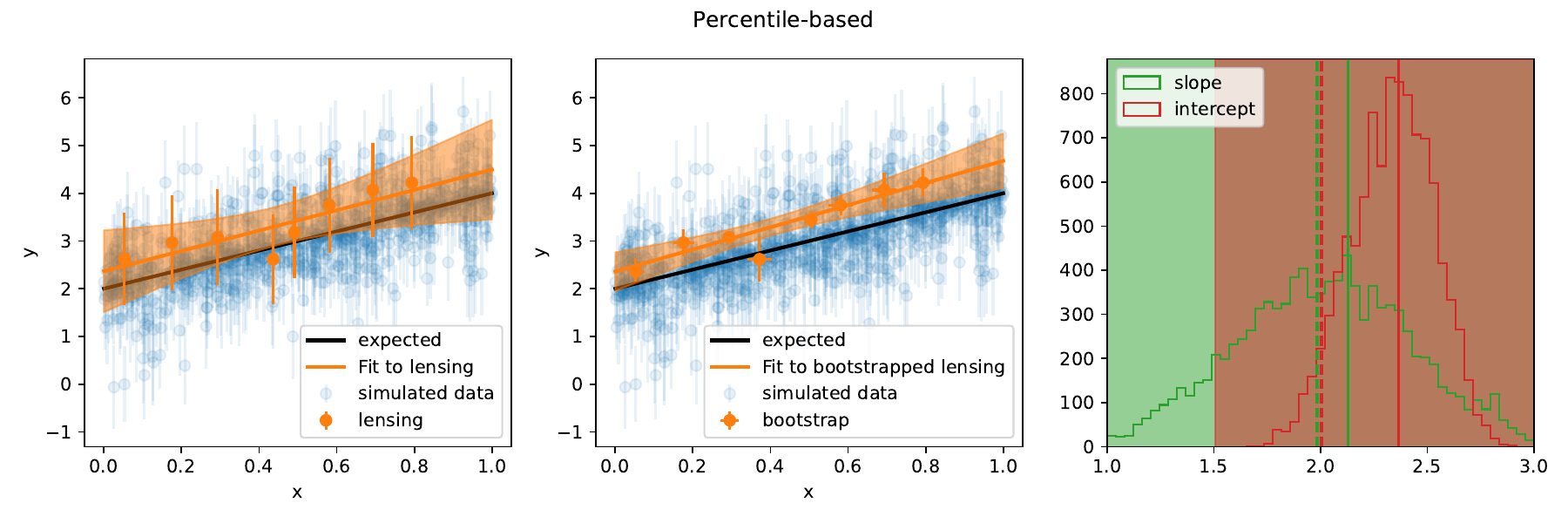}
\includegraphics[width=\linewidth]{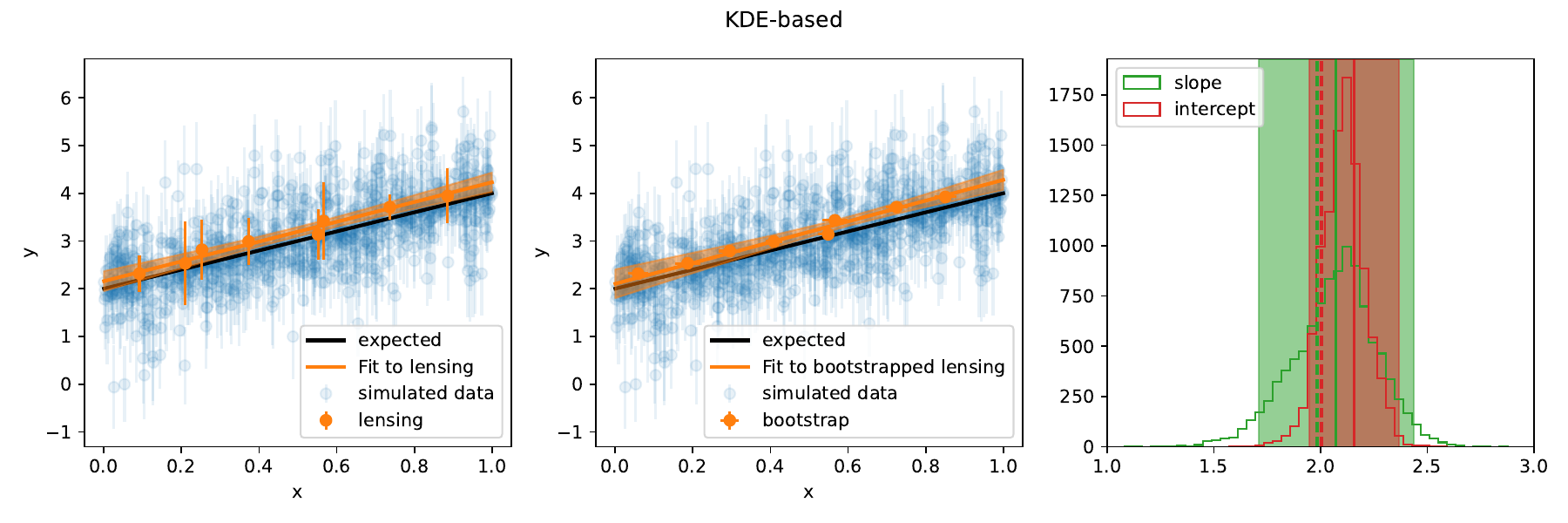}
\includegraphics[width=\linewidth]{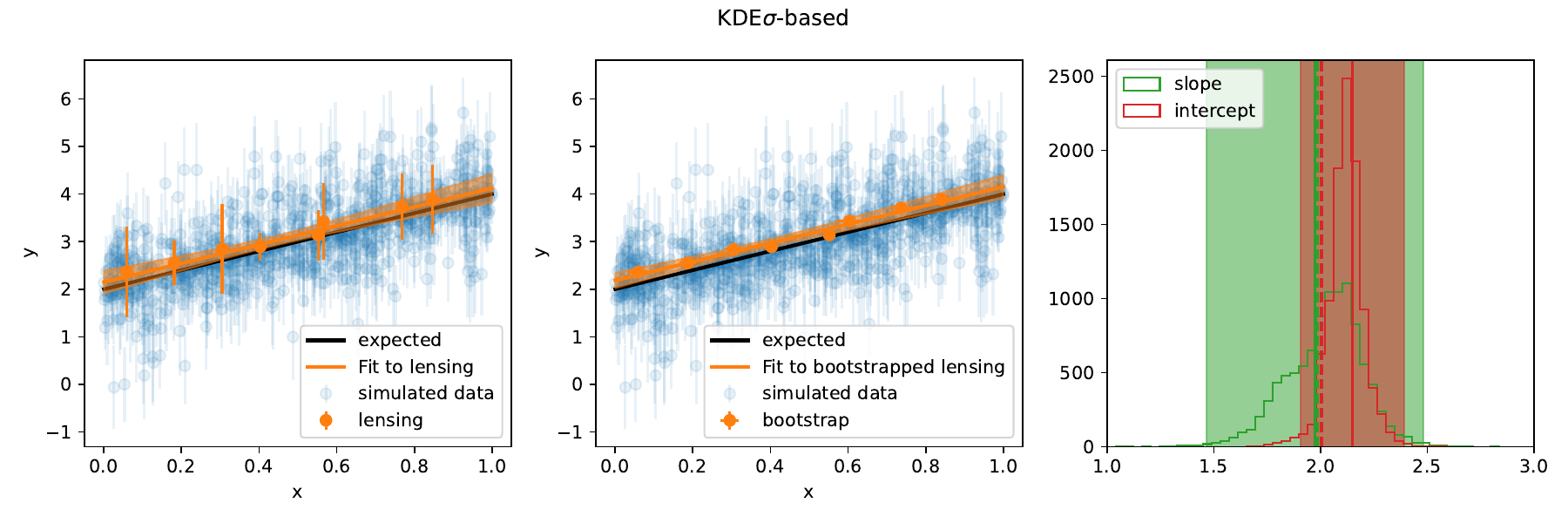}
\caption{Lensing of linear model simulations (blue error bars) using the percentile-based (top row), KDE-based (middle row) and KDE$\sigma$-based methods (bottom row). Orange points show the result of the lensing procedure (left panels) and their boostrapped estimates (middle panels). The right panels show the histograms of the slope (green) and intercept (red) of a linear model fitted to the bootstrap samples as well as the vertical lines marking the prediction intervals, computed from the histograms, where dashed lines are for the best fits and shaded intervals for the prediction intervals. Solid lines represent the non-bootstrapped best-fits.}
    \label{fig:bootstrapSim}
\end{figure*}

\begin{figure*}
    \includegraphics[width=\linewidth]{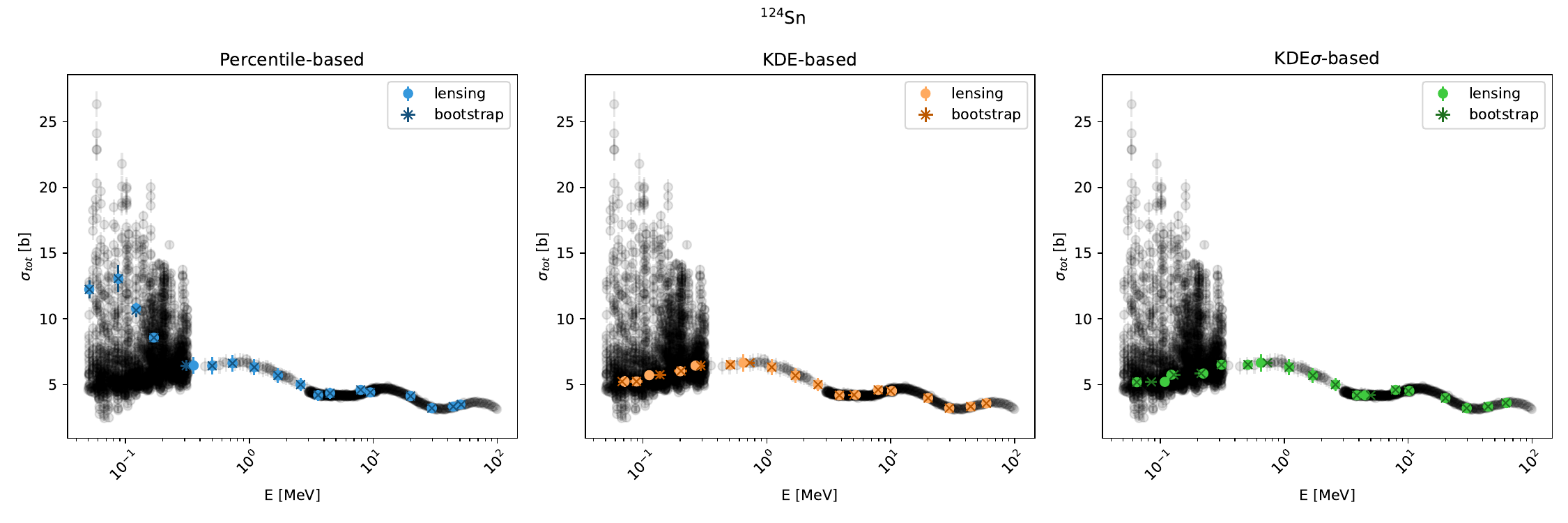}
    \includegraphics[width=\linewidth]{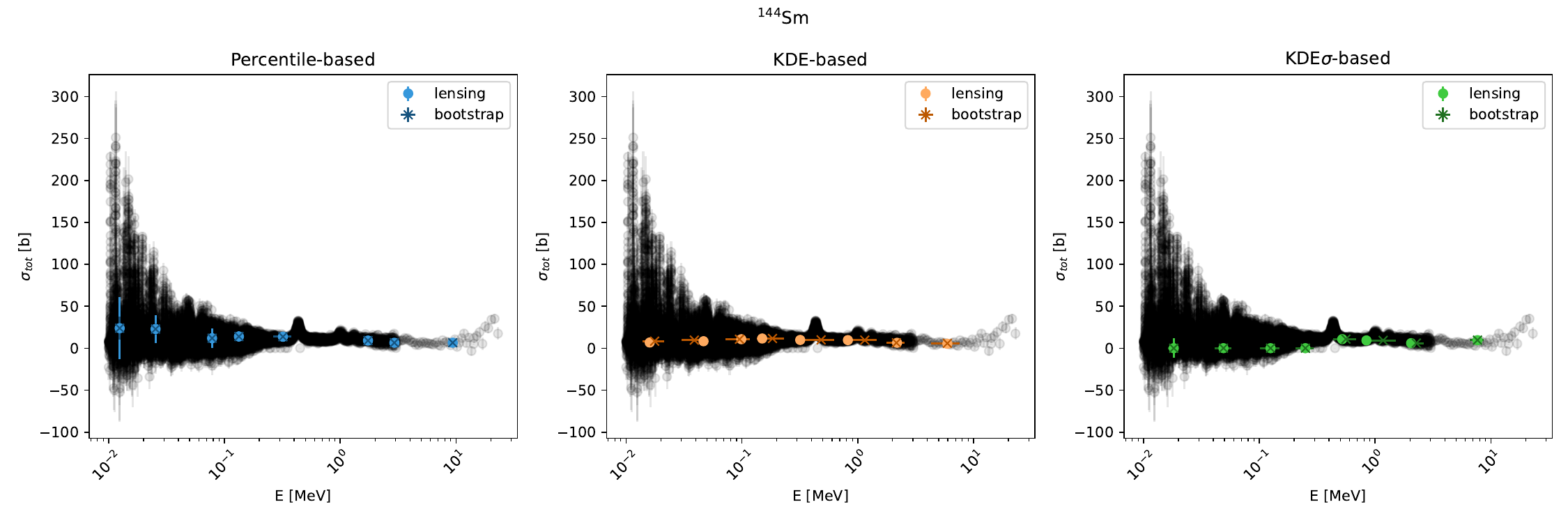}
    \caption{ Total cross section dataset and the lensed subset. The left panels display the percentile-based estimates, the middle panels show the KDE-based estimates, and the right panels present KDE estimates incorporating heteroscedastic errors. Black points represent the full dataset, while blue, orange, and green points indicate the lensed subsets. Crosses denote the bootstrap estimates for the lensing bin values. The data presented is total cross section  for $n+{}^{124}$Sn (top)  and $n+{}^{144}$Sm (bottom).}\label{fig:nuclei}
    \end{figure*}
    \begin{figure*}
    \includegraphics[width=\linewidth]{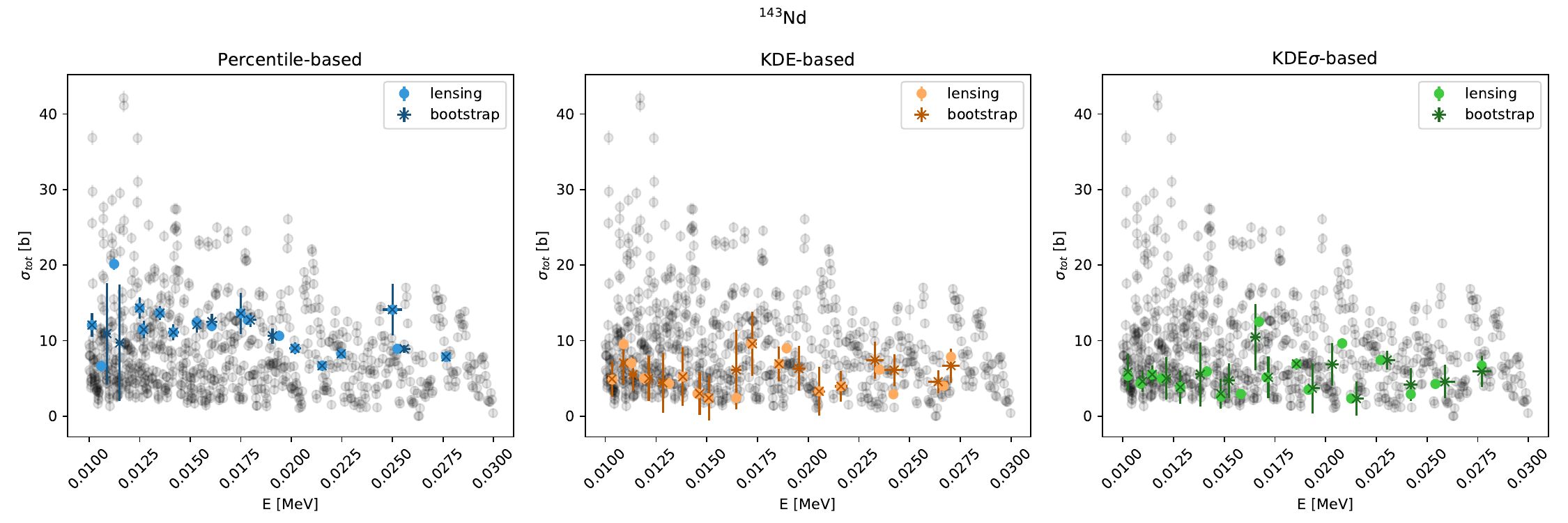}
    \includegraphics[width=\linewidth]{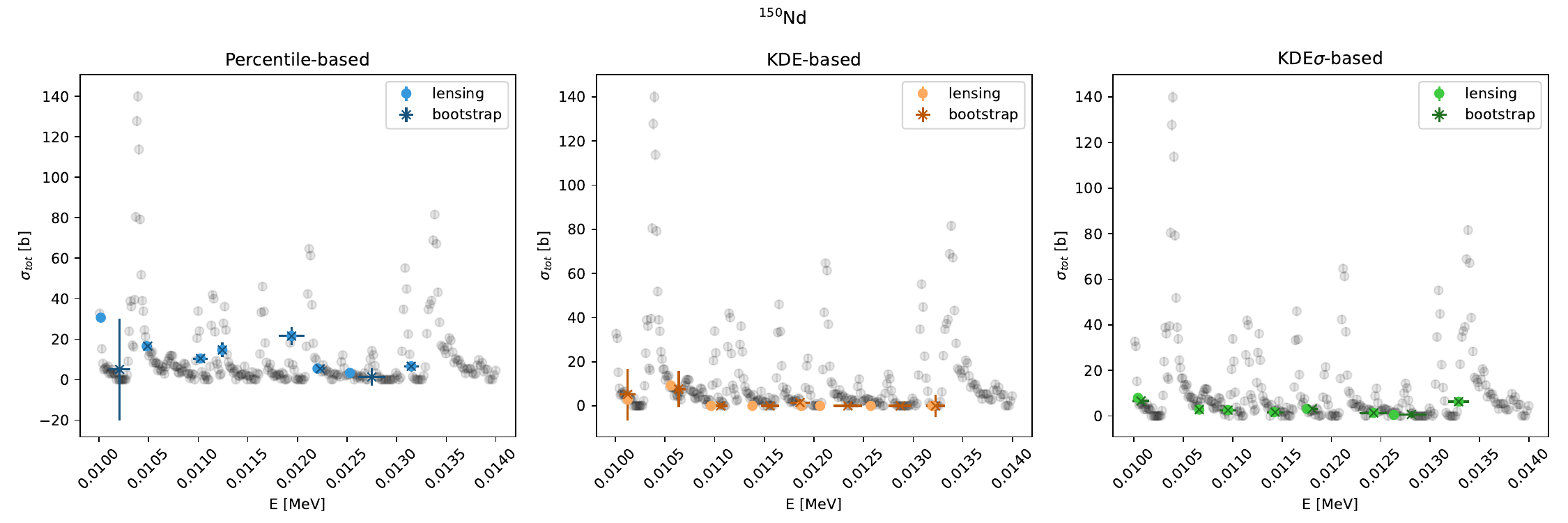}
    \caption{Same as Fig.~\ref{fig:nuclei},  the data presented is total cross section  dataset and the lensed subset for $n+{}^{143}$Nd (top)  and $n+{}^{150}$Nd (bottom).The left panels display the percentile-based estimates, the middle panels show the KDE-based estimates, and the right panels present KDE estimates incorporating heteroscedastic errors. Black points represent the full dataset, while blue, orange, and green points indicate the lensed subsets. Crosses denote the bootstrap estimates for the lensing bin values.  }
    \label{fig:nucleicont}
\end{figure*}
\begin{figure*}
     \centering\
    \begin{subfigure}[b]{0.5\textwidth}\    
        \includegraphics[width=\linewidth,trim={0 0 0 .75cm},clip]{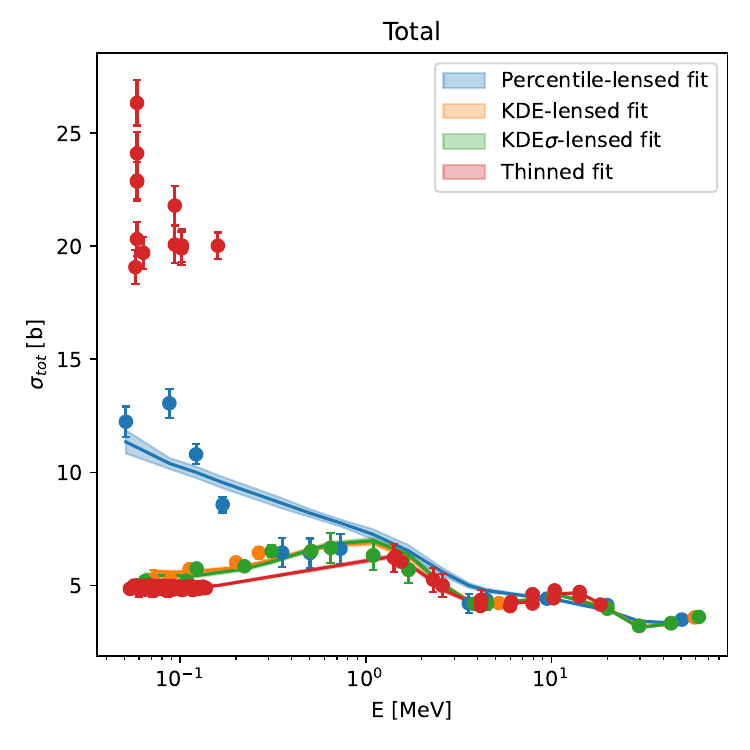}\
        \end{subfigure}\begin{subfigure}[b]{0.5\textwidth}\    
    \includegraphics[width=\linewidth,trim={0 0 0 .75cm},clip]{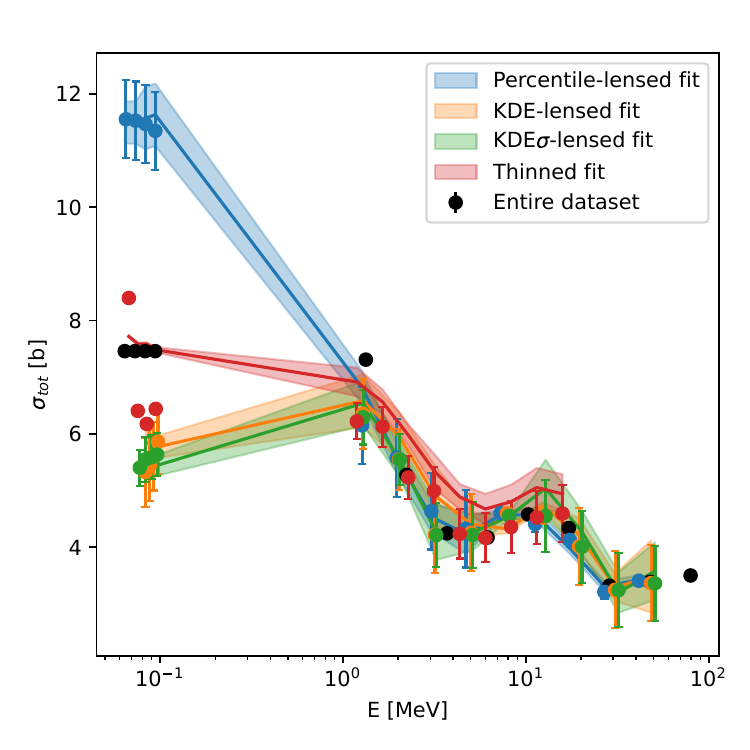}\
    \end{subfigure}\
    \caption{The reduced  ${}^{124}$Sn datasets and the corresponding CoH$_3$ model fits for the total cross section channel. Left panel shows the lensed and thinned data for ${}^{124}$Sn, while the right panel shows the FBET-smoothed lensed data and the FBET-smoothed thinned data. The black dots show the FBET smoothing applied to the entire dataset. }\
    \label{fig:ThinningSn124}\
\end{figure*}
\begin{figure*}
     \centering\
    \begin{subfigure}[b]{0.5\textwidth}\    
        \includegraphics[width=\linewidth,trim={0 0 0 .75cm},clip]{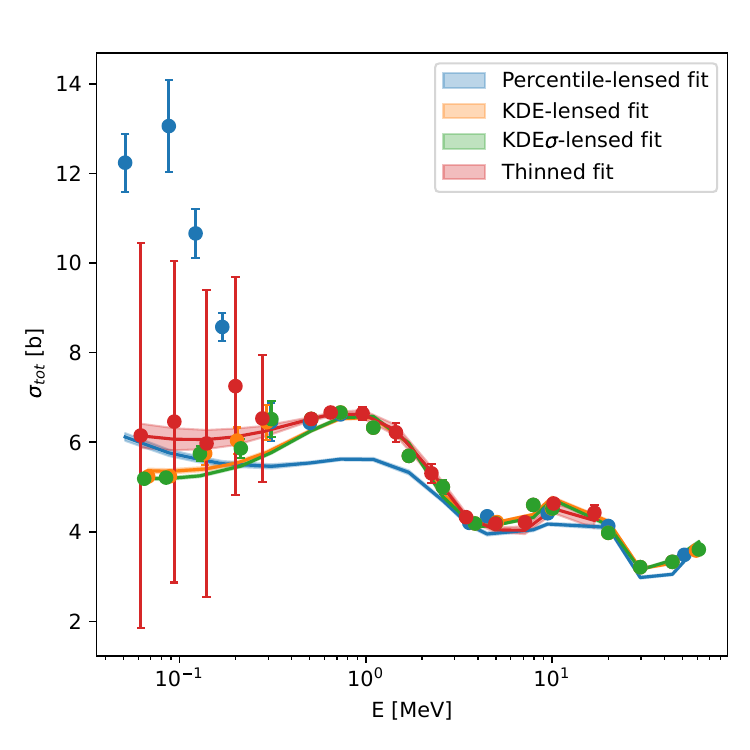}\
        \end{subfigure}\begin{subfigure}[b]{0.5\textwidth}\    
    \includegraphics[width=\linewidth,trim={0 0 0 .75cm},clip]{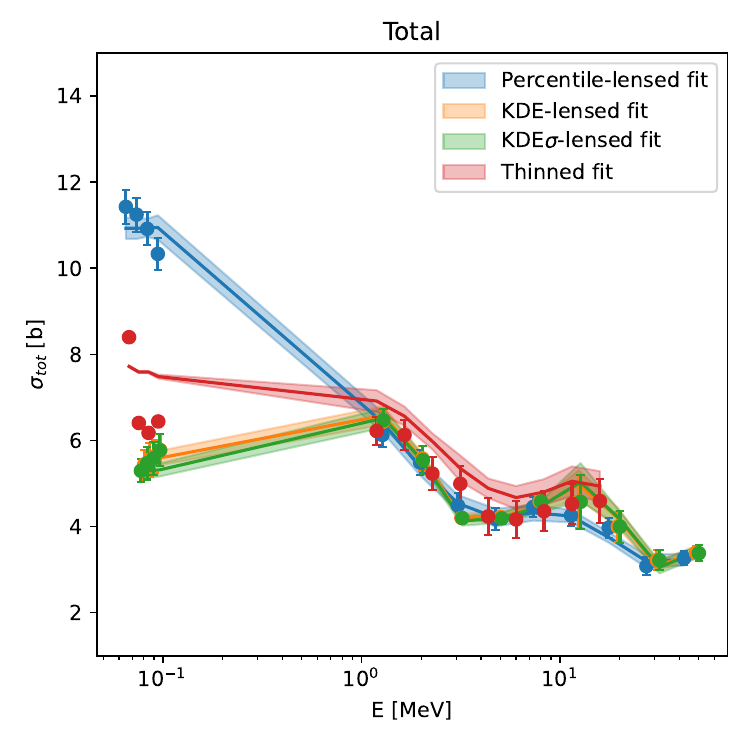}\
    \end{subfigure}\
    \caption{The bootstrapped and reduced  ${}^{124}$Sn datasets and the corresponding CoH$_3$ model fits for the total cross section channel. Left panel shows the lensed and thinned data for ${}^{124}$Sn, while the right panel shows the FBET-smoothed lensed data and the FBET-smoothed thinned data. }\
    \label{fig:ThinningSn124b}\
\end{figure*}

\subsection{Kernel-density-based estimates}\label{sec:kde}
We also developed a modified version of the lensing method from Sec. \ref{sec:percentile} that uses kernel density estimates (KDE) of the bin $b'$
\begin{equation}
    K_h\left(y\mid h,b'\right)=\sum\limits_{ x_j\in[x_{b'},x_{b'+1}]}\frac{e^{-\frac{(y-y_i)^2}{2h^2}}}{\sqrt{2\pi} h n_{b'}},
\end{equation}
where $n_{b'}$ is the number of measurements in the bin $b'$.
In this variant, the index of the measurement from bin $b$ that maximizes the probability density of data from bin $b'$, $L_{b'}$,
\begin{align}
    i_{b,b'} =\argmax\limits_{x_i\in[x_b,x_{b+1}]} K_h\left(y_i\mid h,b'\right).
\end{align}
This method has the advantage of taking into account more details of the measurement values in the bin $b'$ and is the limiting case of using all of the percentiles for computing the likelihood $L_b(y,\sigma_y)$. 

A straightforward generalization of the KDE method involves taking errors $\sigma_y$ instead of a fixed bandwidth $h$.
This $\sigma$-based KDE lensing approach naturally adapts the KDE method to the measurement uncertainty of each data point, thereby preserving the probabilistic structure of the input distribution.
Unlike traditional KDE lensing, which applies a fixed kernel width regardless of the measurement precision, the $\sigma$-based variant ensures that more precise data contribute sharper likelihood features, while noisier data are appropriately smoothed.
This makes the method especially advantageous when the data exhibit heteroscedasticity, i.e., when uncertainties vary significantly across samples.
  \section{Methods}\label{sec:method}
  Our lensing method can be used as a pre-processing step when analyzing dense data, such as nuclear cross-section data in the resonance region. We demonstrate the entire procedure by applying the FBET data smoothing procedure and fitting the CoH$_3$ model. Therefore, in sections \ref{sec:fbet}, \ref{sec:coh}, and \ref{sec:bootstrap} we briefly describe the standard FBET techniques, the CoH$_3$ optical potential model and the bootstrap technique used for testing.
\subsection{FBET}\label{sec:fbet}
In nuclear data evaluation the FBET method is often used to incorporate prior information by a linearized approximation to the Bayes rule \cite[see, e.g.,][for details] {leeb_consistent_2008}. The experimental nuclear cross sections $\mathbf{y}_m$ are modeled as functions of a random variable, $\mathbf{y}$ that describes cross sections evaluated at chosen energy grid points, subject to model $M$, i.e., $\mathbf{y}_m=f_M(\mathbf{y})$. This procedure assumes the multivariate normal distribution for the prior values of $\mathbf{y}$:
\begin{equation}
p(\mathbf{y}\mid M ) = N e^{  -\frac{1}{2} (\mathbf{y} - \mathbf{y}_0)^T A_0^{-1}(\mathbf{y} - \mathbf{y}_0)  },
\end{equation} 
where vector $\mathbf{y}_0$ is the \emph{a priori} mean value of the parameter vector $\mathbf{y}$, $A_0$ is the a-priori covariance matrix of the parameters, and $N$ the normalization factor. By further assuming that experimental data have uncertainties described by the experimental covariance matrix $B$ - typically containing measurement uncertainties on the diagonal - one obtains the posterior distribution 
\begin{align} p(\mathbf{y} \mid M) &= N' \exp\Big(-\frac{1}{2} (\mathbf{y} - \mathbf{y}_0)^T A_0^{-1}(\mathbf{y} - \mathbf{y}_0)\\
&- \frac{1}{2} (f(\mathbf{y}) - \mathbf{y}_m)^T B^{-1} (f_M(\mathbf{y}) - \mathbf{y}_m)\Big), \end{align}
were $N'$ is a new normalization constant that absorbs the contributions from both the prior and likelihood terms.
The procedure introduces the sensitivity matrix to linearize the relationship between $\mathbf{y}_1$ and $f_M(\mathbf{y}_1)$ as $f_M(\mathbf{y}_1) = S \mathbf{y}_1$, resulting in the following linearized estimates for the mean \emph{a posteriori} value $\mathbf{y}_1$, and the corresponding \emph{a posteriori} covariance matrix $A_1$:
\begin{align}
    \mathbf{y}_1 &= \mathbf{y}_0 + A_0S^T (S A_0 S^T + B)^{-1}(\mathbf{y}_m - f_M (\mathbf{y}_0))\\
    A_1&=A_0-A_0 S^T(S A_0 S^T+B)^{-1}S A_0.
\end{align}
The matrix $A_1$ contains correlations between all experimental data, which can then be used in the fitting process.
\subsection{CoH${}_3$ optical model fitting via LM algorithm}\label{sec:coh}
The CoH${}_3$ code computes nuclear reaction cross sections in the fast energy range, including total, shape elastic, and direct inelastic scattering, as well as direct/semidirect capture, pre-equilibrium emission, and particle and $\gamma$-ray emission processes \citep{Kawano2019}. Importantly, CoH${}_3$ computes particle transmission coefficients internally and does not rely on external optical model solvers. For deformed nuclei, the code employs a coupled-channels formalism extended with rotational and vibrational models. Throughout this work, we use the global Koning--Delaroche optical potential \citep{Koning2003}, as implemented in CoH${}_3$, for all neutron-induced reaction calculations.

To determine the effectiveness of our data thinning method, we employed a standard damped Levenberg–Marquardt (LM) algorithm \citep{marquardt1963algorithm} to optimize multiplicative scaling factors (tweaks) of the six optical model parameters of the neutron Koning--Delaroche optical potential \citep{Koning2003}. These scaling factors tweak the real potential depth, $t_{OV}$, the imaginary surface potential depth, $t_{OW}$, the real potential radius, $t_{ORV}$, the imaginary surface potential radius, $t_{ORW}$, the real potential diffuseness, $t_{OAV}$, and the imaginary surface potential diffuseness, $t_{OAW}$.

A schematic overview of the fitting procedure is provided in Fig.~\ref{fig:Schema}. 
The fits were performed with experimental total cross-section data, covering neutron energies from approximately 0.01~MeV to 10~MeV.

The LM algorithm is an iterative nonlinear least-squares optimizer modulated by a varying damping parameter. At each iteration, a proposed parameter update is accepted or rejected based on the reduction of the objective function, defined here as the $\chi^2$ between the experimental data and model predictions.
The fitting procedure uses the residuals $\mathbf{r}(\theta)$ between the FBET-smoothed data $\mathbf{y}_1$ and the model predictions $\mathbf{f}(\theta)$,
\begin{equation}
\mathbf{r}(\theta) = A_1^{-1/2}\left(\mathbf{y}1 - \mathbf{f}(\theta)\right),
\end{equation}
along with the corresponding model Jacobian matrix
\begin{equation}
J = \nabla^T_\theta \mathbf{r}(\theta) = -A_1^{-1/2} \nabla^T_\theta \mathbf{f}(\theta),
\end{equation}
to compute the parameter update step $\delta\theta$ via the standard Gauss–Newton formula \cite{Levenberg1944,marquardt1963algorithm}:
\begin{equation}
\delta\theta = -(J^T J)^{-1} J^T \mathbf{r}.
\end{equation}

To quantify uncertainty in the fitted parameters, we estimate the parameter covariance matrix $\Sigma_\theta$ using the Cramér–Rao bound:
\begin{equation}
\Sigma_\theta = (J^T J)^{-1}.
\end{equation}
This matrix is then propagated to the model output to obtain the cross section covariance matrix,
\begin{equation}
\Sigma_\mathbf{y} = \nabla^T_\theta \mathbf{f}(\theta)  \Sigma_\theta  \nabla_\theta \mathbf{f}^T(\theta).
\end{equation}
The cross section uncertainties on the predicted cross sections, $\sigma_\mathbf{y}$, are defined as the square roots of the diagonal entries of $\Sigma_\mathbf{y}$ and represent one standard deviation around the model prediction. These are used to define confidence intervals on the energy grid: 
\begin{equation}[\mathbf{f}(\theta)-\sigma_\mathbf{y}(\theta),\mathbf{f}(\theta)+\sigma_\mathbf{y}(\theta)].\end{equation}

\subsection{Bootstrap}\label{sec:bootstrap}
To assess the variability and robustness of our estimates, we employ the bootstrap method \citep{Efron1979}, a non-parametric resampling technique that approximates the sampling distribution of a statistic by repeatedly drawing samples with replacement from the observed data. This approach enables estimation of standard errors, confidence intervals, and the stability of fitted parameters in both the percentile- and kernel-density-based methods. 

\subsection{Classical thinning}\label{sec:thinning}
The thinning through estimation method, as described in \cite{Ochotta2005}, iteratively reduces a dataset by removing the least informative points with respect to a chosen estimator. The procedure is designed to preserve predictive power while reducing redundancy in large spatial or multidimensional datasets.
The method starts with a dataset of $N$ pairs of values, $(x_i,y_i)$ indexed by $i\in P_0=\{1,\cdots, N\}$, and a kernel regression estimator $\hat{y}(x,S)$, that takes a subset $S\subset P_0$ and evaluates a chosen kernel $K_h$ for a particular point, $x$:
\begin{align}
    \hat{y}(x,S) = \frac{\sum\limits_{j \in S} K_h(x, x_j) y_j}{\sum\limits_{j \in S} K_h(x, x_j)},
\end{align}
where $K_h$ is the Gaussian kernel symmetric kernel function with bandwidth $h$:
\begin{align}
    K_h(x, x') = \exp\left(-\frac{\|x - x'\|^2}{2h^2}\right).
\end{align}
For each point $i$ the associated mean squared error, $e(x,S)$ is computed
\begin{align}
    e(x_i,S) = \left(y_i - \hat{y}(x_i,S)\right)^2.
\end{align}
The algorithm computes $e(x_i,P_0\setminus\{i\})$ for all $i\in P_0$ and proceeds by ranking the points based on $e(x_i,P_0\setminus\{i\})$ for all $i\in P_0$ and removes the point with the smallest error, yielding a subset $P_1$. The procedure is then iterated, for a step $k$ the expression can be written as 

\begin{equation}
    P_{k+1}=P_{k}\setminus \argmin_{i\in P_k}\left(e(x_i,P_k\setminus\{i\})\right)
\end{equation}

This process continues until a desired target mean squared error is reached. The resulting reduced dataset $P_k$ retains the key predictive characteristics of $P_0$ while substantially reducing its size.

\section{Testing on simulated data}\label{sec:testing}

We demonstrate our method on simulated data taken from a normal distribution, following a linear model with heteroscedastic, uncorrelated errors.  To this end, we constructed a random sample of a linear model with set values of slope and intercept, computed the fit of a linear model to the lensed subsample, and estimated the reliability of these fits using bootstrap sampling of the simulated dataset.

The data are simulated using uniform distributions for a variable, $x$, and its standard deviation, $\sigma$. Samples of a dependent variable, $y$, are then drawn from a normal distribution.
\begin{align}
     x&\sim U(0,1)\\
     \sigma&\sim U(.01,1)\\
     y(x,\sigma)&\sim \mathcal{N}(2x+2,\sigma^2\mid x,\sigma).
\end{align}
For this example a simulated sample size of $N_s=1000$. 

The results of fitting a linear model to the lensed subsets of the simulated data are shown in the left panels of Fig. \ref{fig:bootstrapSim}. In Table \ref{tab:bootstrap}, we show the comparison of the lensing methods based on the t-tests on the slope and intercept, computed with respect to the expected slope and intercept values, using the simulated relation $y=2x+2$. 

To evaluate the stability and reliability of the lensing procedure, we conducted bootstrap subsampling $N_B=10000$ times. In the middle panels of Fig.~\ref{fig:bootstrapSim}, we show the bootstrapped values - the median and standard deviation of the bootstrap subsamples for each $x$-bin. The orange line is a fit to the orange points. 

We then fit each bootstrap of the lensed subsample with a linear model and compare the resulting slope and intercept distributions in Fig.~\ref{fig:bootstrapSim}. We find that the prediction intervals ($\sqrt{N_s+1}\times$ estimated parameter error of the simulated sample, and $\sqrt{N_B+1}\times$ estimated parameter error of the lensed subsample), shown as vertical lines, agree well with the bootstrap distributions.

The Chow $F$ statistics \cite{Chow1960} and their corresponding P-values were computed to compare the best-fitting model's evaluation to the best-fitting evaluations of the bootstraps. In brief, the method computes the $F$-statistic based on the sum of squares of the residuals of the best-fitting model's residual sum of squares, $RSS_m$, the bootstrap best-fitting model's residual sum of squares $RSS_b$ and the residual sum of squares of a union of the lensing dataset and the bootstrapped lensing dataset, $RSS_u$ as
\begin{equation}
F = \frac{(RSS_u - (RSS_m + RSS_b)) / k}{(RSS_m + RSS_b)/(n_m + n_b - 2k)},
\end{equation}
where $k$ is the number of model parameters, i.e. $k=2$, and $n_m$ and $n_b$ are the sizes of the lensed dataset and its bootstrap, respectively. We list the Chow $F$ statistic values and the corresponding p-values (using the $F(k,n_m + n_b - 2k)$ distribution) in table \ref{tab:bootstrap}. We find that there is no statistically significant difference between the lensed dataset and its bootstrap ($P>0.1$). The method therefore gives reliable results for all three lensing variants.

Computation times for the lensing methods are shown in Table~\ref{tab:bootstrap}, reported as the median and standard deviation over bootstrap simulations. We find that the percentile-based lensing method is significantly faster than the KDE-based approach. For comparison, the classical thinning method required over 5 minutes to process the same dataset, due to the higher computational cost at each iteration.

\begin{table*}[]
    \centering
    \begin{tabular}{c c c c c}
\toprule
Lensing method & ${}^{124}$Sn  & ${}^{150}$Nd &   ${}^{143}$Nd& ${}^{144}$Sm\\
\midrule
Percentile-based & $0.0312\pm 0.0014$ & $0.01268\pm 0.00091$& $0.0146\pm 0.0013$ & $0.0121\pm0.0023$\\
KDE-based &$0.237\pm 0.018$ & $0.0898\pm 0.0078$& $0.1218\pm 0.0044$ & $0.546\pm 0.013$\\
KDE$\sigma$-based & $0.196\pm 0.016$ & $0.0715\pm 0.0064$& $0.1167\pm 0.0035$  & $0.446\pm 0.015$\\
Thinning & $1500\pm 390$ & $9.0\pm 1.1$& $133.5\pm 7.0$ & $1.98\times 10^{5}$\footnote{Evaluation for the ${}^{144}$Sm thinning method was omitted due to long runtimes.}\\
\bottomrule
\end{tabular}
\caption{Comparison of computation times (in seconds) for the lensing methods and the thinning method across the selected nuclei. Uncertainties were estimated using 100 bootstrap samples. }
    \label{tab:timings}
\end{table*}
\section{Results}\label{sec:results}

We demonstrate our method on total cross section data for four neutron-induced reactions—n+${}^{124}$Sn \cite{Carlton1996,Harper1982,Musaelyan1989,Pruitt2020,Rapaport1980,Dukarevich1967}, n+${}^{144}$Sm \cite{Macklin1993,Dyumin1973}, n+${}^{143}$Nd \cite{Tellier1971,Wisshak1998}, and n+${}^{150}$Nd \cite{Tellier1971}—chosen to provide a diverse sample that includes both resonance-rich and smooth, non-resonant measurements. These results are intended to showcase the efficacy of our data thinning approach; a comprehensive multi-channel evaluation \cite{Imbrisak2025b} and extensions for highly unbalanced datasets \cite{Imbrisak2025a} will be presented in follow-up work.

Figs.~\ref{fig:nuclei} and \ref{fig:nucleicont} display the full datasets alongside the subsets extracted using our lensing algorithm. Left panels show percentile-based estimates, middle panels show KDE results, and right panels depict KDE estimates with heteroscedastic error treatment. In all panels, black points denote the complete dataset; colored points (blue, orange, green) represent the lensed subsets; and crosses indicate bootstrap-based estimates for the lensing-bin values.

The selected datasets span both narrow and wide energy ranges, allowing us to evaluate the robustness of the thinning procedure across different experimental regimes. In particular, ${}^{143}$Nd and ${}^{150}$Nd offer dense low-energy data, suitable for benchmarking statistical consistency, but lack high-energy measurements (above $\sim$1 MeV) required for meaningful CoH$_3$ model fits. In contrast, ${}^{124}$Sn and ${}^{144}$Sm datasets extend into the fast neutron regime and enable full model optimization. To illustrate this difference, we perform fits for ${}^{124}$Sn (Sec.~\ref{sec:Sn124}).

Our results indicate that the lensing method tends to exclude points within pronounced resonance peaks, especially in the ${}^{124}$Sn and ${}^{144}$Sm cases, thereby favoring smoother samples more representative of the average cross section. However, for reactions with limited energy ranges (notably ${}^{143}$Nd and ${}^{150}$Nd), the percentile-based bootstrap estimates occasionally show offset from the lensed subsets, likely due to uneven data distribution in the energy–cross-section plane. KDE-based estimates, which take local data density into account, help mitigate this bias.

Table~\ref{tab:timings} reports computation times for the lensing methods, including standard errors from 100 bootstrap simulations. The percentile-based variant is substantially faster than KDE-based lensing. All of the lensing methods are faster than the thinning method. The number of bootstrap samples was chosen to strike a balance between reliable uncertainty estimation and total computation times for bootstrap simulations. 

\subsection*{Application to model fitting to $n+{}^{124}$Sn}\label{sec:Sn124}
To further assess the impact of our data reduction techniques, we applied the FBET data assimilation approach to the $n+{}^{124}$Sn lensed subset (see Fig. \ref{fig:nuclei}), followed by a fit of a 6-parameter CoH${}_3$ model. Figure \ref{fig:ThinningSn124} contrasts the FBET-smoothed CoH${}_3$ fit from the lensed dataset (blue points) with the outcome of the standard data thinning method, which was applied to the full initial dataset before undergoing FBET processing. The comparison reveals that the lensing approach yields estimates with closer agreement to the FBET-smoothed expectations than does the traditional thinning procedure. Notably, while the standard thinning method selects points based on clustering of the most-distant (influential) measurements, our lensing algorithm selectively eliminates resonance-associated points, thereby enhancing the fidelity of the reduced dataset. 

The left panel of Fig. \ref{fig:ThinningSn124} illustrates the raw lensed and thinned data prior to FBET smoothing. These results underscore that both thinning and lensing methods serve best as pre-processing steps in tandem with FBET for subsequent CoH${}_3$ fits. 
In Fig. \ref{fig:ThinningSn124}, we also present the results of applying the FBET procedure to the lensing and thinning reduced total cross section dataset. Notably, the thinning procedure (red points) exhibits a strong correspondence with the FBET-smoothed complete dataset. This agreement stems from the fact that thinning preferentially selects points that lie far from the mean—predominantly capturing the resonance peaks. By contrast, the lensing methods consider the overall distribution of data within an energy bin, thus weighting points closer to the mean cross section. This fundamental difference in data selection explains the observed discrepancies between the two approaches. 

 By condensing the dataset prior to applying FBET, our approach markedly decreases the computational burden traditionally associated with smoothing and parameter fitting, thereby expediting the overall analysis workflow and enabling more rapid exploration of parameter spaces in large-scale nuclear data applications.

In Fig. \ref{fig:ThinningSn124b}, we present the bootstrap estimates for the reduced datasets shown in Fig. \ref{fig:ThinningSn124} and fit these estimates with the CoH${}_3$ model. This analysis is not intended as an alternative to the lensing methods; rather, it illustrates that the bootstrap estimates derived from the lensing approaches yield results consistent with the expected lensed datasets, even after model fitting. These bootstrap estimates are not actual measurements, and both the bootstrap and smoothed bootstrap fits should be regarded as purely illustrative. This can be seen in the behavior of thinning bootstraps (red points in Fig.~\ref{fig:ThinningSn124b}) whose medians, in contrast to the actual thinned subsets in Fig.~\ref{fig:ThinningSn124}, below $1\,\mathrm{MeV}$ are aligned with the lensing methods only because the thinning bootstrap estimates appear symmetrically above and below lensing bootstraps, with significantly larger uncertainties. We find that our method provides robust results since the FBET smoothed bootstraps in the right panel of Fig.~\ref{fig:ThinningSn124b} are in agreement with the non-bootstrapped smoothed data in the right panel of Fig.~\ref{fig:ThinningSn124}.




\section{Conclusion\label{sec:conclusion}}
We have presented a novel data thinning algorithm inspired by optical lensing, designed to improve the pre-processing of nuclear cross section data for subsequent statistical analysis. Our approach prioritizes the selection of structurally informative data points, particularly in resonance-rich regions, and is compatible with standard Bayesian assimilation workflows such as FBET.

By applying our method to both simulated and experimental nuclear data, we have shown that the optical lensing method offers a practical compromise between information retention and computational efficiency. In cases where full dataset processing is prohibitively expensive, our method allows analysts to work with a representative subset that yields statistically comparable outcomes.

While our main focus has been on nuclear cross sections, the generality of the approach suggests broader applicability to other fields that involve dense experimental datasets with correlated structures.

\acknowledgments
This work was performed under the auspice of the U.S. Department of Energy by Los Alamos National Laboratory under 89233218CNA000001. Research reported in this publication was supported by the U.S. Department of Energy LDRD program at Los Alamos National Laboratory (Project No. 20240004DR).
\bibliography{references.bib}
\newpage

\end{document}